# Disentangling the Roles of Dissolved Oxygen, Common Salts, and pH on the Spontaneous Hydrogen Peroxide Production in Water: No $O_2$, No $H_2O_2$

Muzzamil Ahmad Eatoo[1,2,*] & Himanshu Mishra[1,2,*]

[1] Environmental Science and Engineering (EnSE) Program, Biological and Environmental Science and Engineering (BESE) Division, King Abdullah University of Science and Technology (KAUST), Thuwal, 23955-6900, Kingdom of Saudi Arabia

[2] Interfacial Lab (iLab), King Abdullah University of Science and Technology (KAUST), Thuwal 23955-6900, Saudi Arabia

*Correspondence: muzzamil.eatoo@kaust.edu.sa and himanshu.mishra@kaust.edu.sa

**TOC Image**

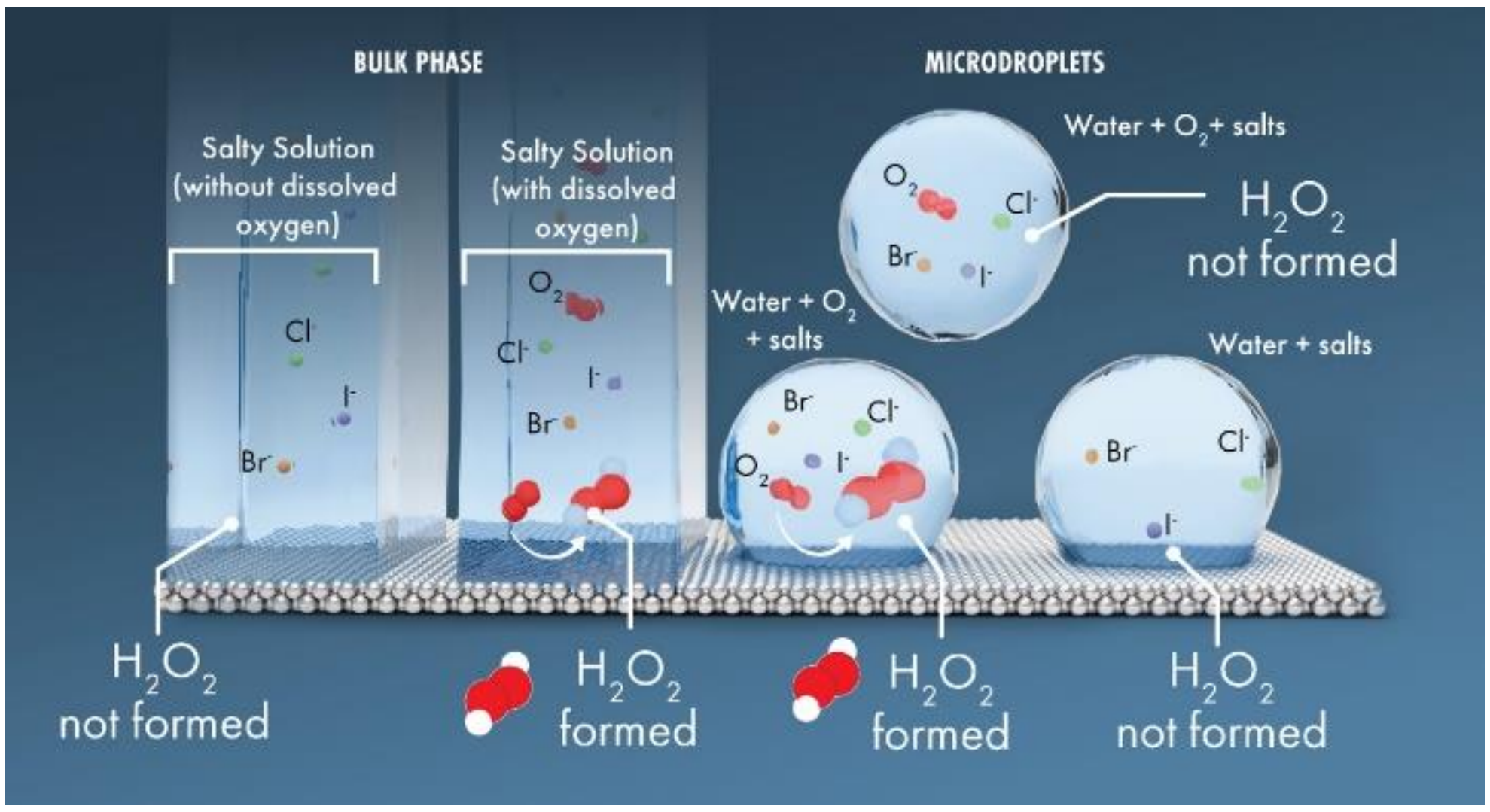
BULK PHASE
MICRODROPLETS
Salty Solution (without dissolved oxygen)
Salty Solution (with dissolved oxygen)
Water + $O_2$ + salts
Water + $O_2$ + salts
Water + salts
$O_2$
$Cl^-$
$Br^-$
$I^-$
$H_2O_2$ not formed
$H_2O_2$ formed
$H_2O_2$ formed
$H_2O_2$ not formed
$H_2O_2$ not formed

**Abstract:**

Despite the mounting evidence proving that the air–water interface or the microdroplet geometry has nothing to do with the spontaneous formation of hydrogen peroxide ($H_2O_2$), the myth persists. Three recent studies by George & co-workers give credence to the myth by showing connections between the spontaneous formation of hydroxyl ($HO^\bullet$) radicals and hydrogen peroxide ($H_2O_2$) in sprayed microdroplets with the solution pH, dissolved salts, nebulizing gas, and the gaseous environment. They report that among halides ($Cl^-$, $Br^-$, and $I^-$), $Br^-$ dominates the $H_2O_2$ formation because of its ability to donate electrons. Also, they conclude that the $H_2O_2$ production at the air–water interface scales with water's alkalinity. In response, we apply a broad set of techniques, spanning $^1$H-NMR, potentiodynamic polarization, electron microscopy, and hydrogen peroxide assay kit (HPAK) fluorometry, to reexamine these claims. Our experiments reveal that regardless of the halide present in water, the air–water interface or the microdroplet geometry does not drive the $H_2O_2$ formation. In fact, it is the reduction of $O_2$ at the solid–water interface, that produces $H_2O_2$, i.e., in the absence of $O_2$, no $H_2O_2$ is formed regardless of the halide. We explain the relative dependence of $H_2O_2$ concentrations on the halides based on their propensity to drive pitting corrosion ($Cl^- > Br^- > I^-$). As the pits appear in the passivating layer, exposing the metal, $H_2O_2$ is consumed in further oxidation. Next, we disprove the claim of alkalinity-driven $H_2O_2$ formation by demonstrating that aluminum and titanium surfaces more $H_2O_2$ in acidic and alkaline conditions, respectively. Taken together, these findings refute the conclusions of George & co-workers and others regarding spontaneous $H_2O_2$ generation at the air-water interface. The following mnemonic captures our conclusion: no $O_2$, no $H_2O_2$.

## Introduction:

The spontaneous formation of $H_2O_2$ in water microdroplets, without a catalyst or external energy, has sparked one of the most prominent debates in the chemical science community. Zare & co-workers have claimed that water microdroplets (<20 $\mu$m diameter) formed via pneumatic spraying or condensation spontaneously generate hydrogen peroxide ($H_2O_2$); albeit, the reported quantities have varied from 180–0.3 μM in the last 6 years[1-5]. They have maintained that ultrahigh electric field fluctuations at the air–water interface are responsible for this chemical transformation, basing on the studies led by Mundy[6], Amotz[7], Head-Gordon[8], and Min[9]. Contrary to the mechanism suggested by Head-Gordon & co-workers, Colussi[10, 11] has put forth a thermochemistry-based argument that entails dehydration of ions at the air–water interface driving the $H_2O_2$ formation, and Cooks[12] has been suggested that the water radical cation is responsible for this chemistry. Is this a hitherto unrealized anomalous property of water?

Our group has challenged these claims and conclusions by demonstrating that the microdroplet geometry or the air–water interface, and, therefore, the ultrahigh instantaneous electric fields therein, do not generate $H_2O_2$.[13-16] Potential artifacts arising from the use of ultrasonication while creating microdroplets[16] and ambient contamination[15] have been pointed out. As well, we have demonstrated that small quantities of $H_2O_2$ (~1 μM) may be formed spontaneously at water–solid interfaces, implicated in these experiments, due to the reduction of dissolved oxygen gas ($O_2$)[14]. As a corollary, we have demonstrated that if $O_2$ is eliminated from the water, then $H_2O_2$ is not formed in the sprays or at the solid–water interface down to a 50 nM detection limit[14]. Recently, Koppenol & co-workers[17], Ruiz-Lopez and co-workers[18], Hassanali & co-workers[19], Williams & co-workers[20, 21], and Rodríguez-López & co-workers[22], have corroborated our findings through complementary approaches based on thermodynamics, quantum chemistry, mass spectrometry, electrochemistry, and electron spin resonance. For additional resources, see refs.[5, 23-28].

Despite the evidence, the myth of $H_2O_2$ generation in water microdroplets persists. A recent experimental report from George & coworkers gives credence to the electric field argument[29]. They also investigated the effects of sodium salts (NaX | X: $Cl^-$, $Br^-$, or $I^-$) on the $H_2O_2$ production in sprayed microdroplets[30]. They conclude that $H_2O_2$ forms specifically at the air–water interface under the effect of ultrahigh electric fields via the following two mechanisms:(i) $OH^-$ ions undergo charge separation to form $HO^•$ and $e^-$, and $HO^•$ radicals form $H_2O_2$, or (ii) $Br^-$ ions undergo charge separation to form $Br^•$ and $e^-$; next, dissolved oxygen ($O_2$) reacts with $e^-$ and $H^+$ to form $HO_2$, which reacts with $H^•$ (produced from the reaction between $H^+ + e^-$) to form $H_2O_2$. Their experiments reveal that at concentrations below 1mM, $Cl^-$ ions generate the highest $H_2O_2$ concentration in the sprays, while at concentrations above 1mM $Br^-$ ions dominate the generation of $H_2O_2$. However, they conclude that "*only $Br^-$ contributes to the interfacial $H_2O_2$ formation, promoting the production by acting as an electron donor*".[30] Lastly, they reported that in the presence of NaCl or $Na_2SO_4$, the $H_2O_2$ production increases by approximately 40% under alkaline conditions (pH 9) compared to acidic conditions (pH 4)[31]. They attribute all these observations to be arising at the air–water interface, where $OH^-$ and/or $Br^-$ ions drive the $H_2O_2$ formation.

We consider that there are several actors in the $H_2O_2$ story – water, its intrinsic ions, $O_2$, halides, the air–water interface, and various solid–water interfaces – which complicate the interpretation of the results. The last actor represents the incorrigible solid–water interfaces that come into play at various stages of the experiments, viz., during sample preparation (e.g., beakers and glass vials), spraying (e.g., tubing comprised of insulators or metals or alloys) or substrate for condensation, analyte collection (e.g., glassware), transferring (e.g., pipettes), and analytical measurement (e.g., vials, cuvettes, or NMR tubes). So, even if one is forming sprays, it must be checked whether the $H_2O_2$ formation takes place prior to spraying or after spraying. Here, we re-examine the claims of George & co-workers[29-31] by probing $H_2O_2$ formation due to $O_2$ reduction and surface oxidation of Al and Ti as representative materials as a function of halides and the water pH. The following interrelated questions are addressed:

1. Is it possible to spontaneously generate $H_2O_2$ at the air–water interface or the solid–water interface in the absence of $O_2$? In this scenario, do halides play any role?
2. Conversely, what is the influence of halides on the $H_2O_2$ formation in the presence of $O_2$, such as water in equilibrium with the air? In this scenario, where does the $H_2O_2$ formation take place: the air–water interface or the solid–water interface?
3. In the experiments of George & co-workers[30], why did microdroplets containing $Br^-$ at > 1 mM concentration generate higher $H_2O_2$ than those containing $Cl^-$ or $I^-$? Is it due to $Br^-$ ions' ability to undergo "charge separation" to form $Br^• + e^-$ at the air–water interface? If not, could this be explained based on solid surface oxidation?

4. At concentrations < 1 mM, why did the $H_2O_2$ formation in salty microdroplets follow the order $Cl^-$ > $Br^-$ (See Fig.2 in Ref.[30])? What is the role of solid–water interfaces?
5. Does $H_2O_2$ formation always increase with the water pH, as claimed by George & co-workers[32]? Could you present a counter case?

While our experiments reproduce the observations of George & co-workers, we contest their conclusions and provide alternative explanations. Crucially, the $H_2O_2$ generation takes place at the water–solid interfaces due to the reduction of $O_2$, i.e., the air-water interface or the microdroplet geometry has no bearing on this chemical transformation.

## Results

### Effects of halides on the spontaneous formation of $H_2O_2$ in microdroplets

Firstly, $H_2O_2$ formation was quantified in pneumatic sprays formed with 10 mM salt solutions of $Na^+X^-$, where X: Cl, Br, or I, and compared with that in the sprays formed with bulk water (Figure 1a and Methods). Prior to spraying, the bulk solutions were divided into two batches: (i) those containing $O_2$ as per the Henry's law (~4 mg/l at normal temperature and pressure)[33], and (ii) solutions with negligible $O_2$, achieved by boiling, followed by $N_2$(g) bubbling for 45 minutes and keeping them inside an $N_2$-purged container (Methods). Note: the latter batch had an $O_2$ concentration < 0.01 mg/l, as measured via an $O_2$ sensor (WTW Multi 3320). In a clean glovebox filled with $N_2$ gas, microdroplets were generated by shearing streams of abovementioned bulk solutions – flowing at the rate of 25µL/min through a 0.10 mm-wide silica capillary – through pressurized $N_2$(g) gas (100 psi) flowing through an outer concentric tube of 0.43 mm in diameter (Figures 1a-b & S1). Inside the $N_2$ environment, sprayed microdroplets were collected in clean glassware and transferred for analytical measurement. The quantification of $H_2O_2$ was done by two methods, namely (1) $^1$H-NMR using a Bruker 950 MHz Avance Neo NMR spectrometer equipped with a 5 mm *Z*-axis gradient TCI cryoprobe at the temperature of 275 K (50,000 scans with a 50-ms acquisition time and a 1 ms recycle delay between scans), following the protocol of Bax and co-workers[34], and (2) Hydrogen Peroxide Assay Kit (HPAK) for $H_2O_2$ quantification (Methods). The limits of detection were ≥ 50 nM and ≥ 250 nM for $^1$H-NMR and HPAK, respectively. While NMR guaranteed an unambiguous fingerprinting of $H_2O_2$, it took ~2 hrs for scanning time per sample; on the other hand, HPAK offered speed and ease but may be vulnerable to false positives due to chemical interference.

To get the best from either scenario, we cross-validated all the $H_2O_2$ measurements with both techniques and found them to be consistent.

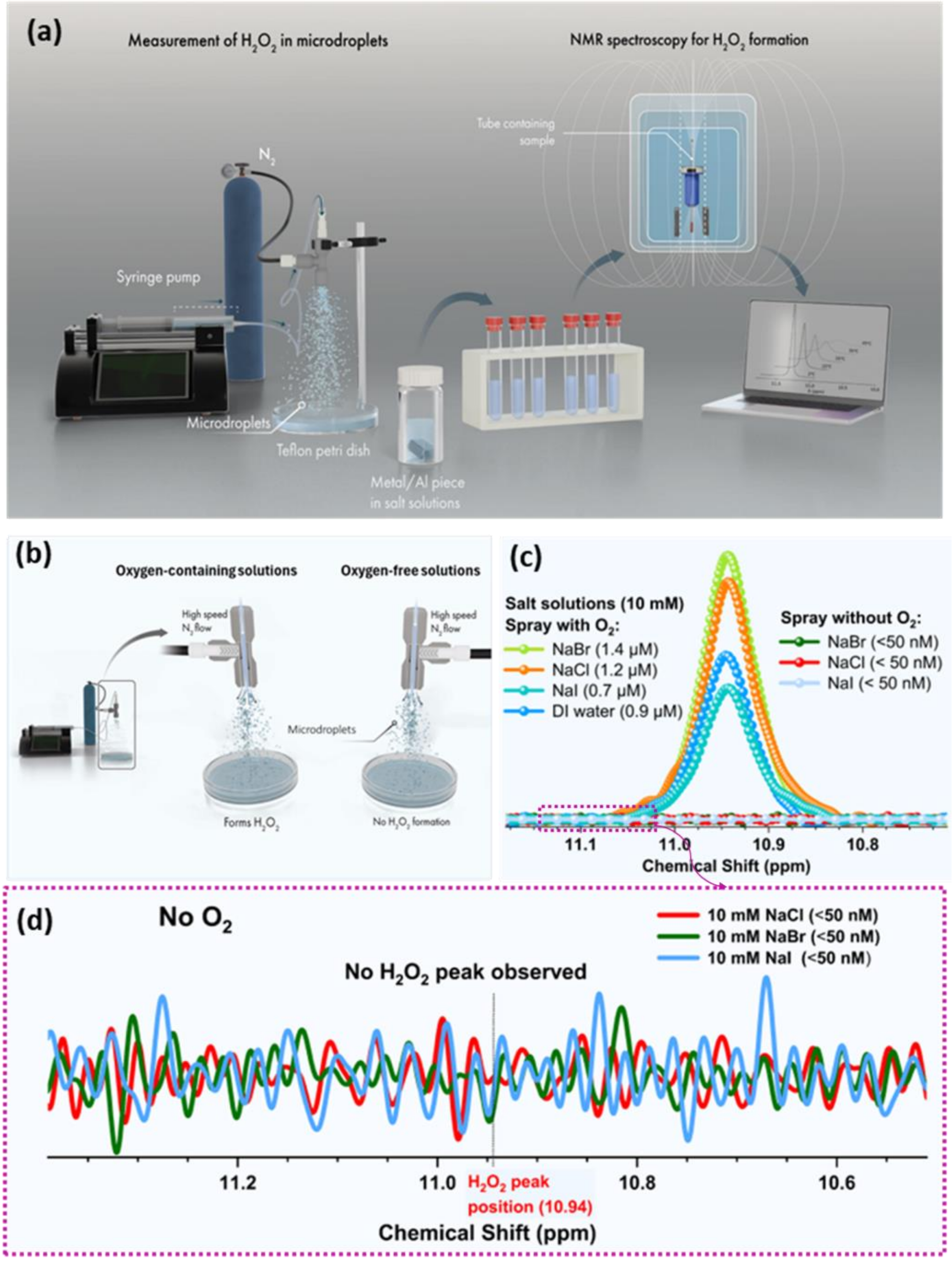

**Figure 1.** (a) Illustration of the experimental setup showing the complete path of $H_2O_2$ investigations in microdroplets by forming microdroplets using high-pressure dry $N_2$ gas and then analyzing the samples using $^1$H-NMR spectroscopy. (b) Illustration of the spray setup to nebulize aqueous streams with and without $O_2$ using high-pressure $N_2$ gas to form microdroplets. The roles of dissolved oxygen ($O_2$) and 10 mM salt ($Na^+X^-$, where X: Cl, Br, or I) concentration on the $H_2O_2$ formation were probed via $^1$H-NMR. (c) Results reveal that in the absence of $O_2$, no $H_2O_2$ was formed (detection limit ≥ 50 nM) regardless of the salt. In the presence of $O_2$ (i.e., solutions saturated with the air), the formation of $H_2O_2$ varied with the salt type. Note: these samples were also analyzed by HPAK, and the results are presented in Figure S2. (d) 1H-NMR results of oxygen-free $N_2$-sprayed salty microdroplets confirming no $H_2O_2$ formation.

The experimental results revealed that for the solutions devoid of $O_2$, sprayed microdroplets did not contain any detectable $H_2O_2$, regardless of the salt (Note: the limit of detection was ≥ 50 nM) (Figure 1c-d). **This is smoking gun evidence that the air–water interface or the microdroplet geometry or the ultrahigh instantaneous electric fields do not form $H_2O_2$,** contradicting several proposed mechanisms[8, 10-12]. Notably, in the absence of $O_2$, solid–water interfaces also do not form $H_2O_2$. No wonder, George & coworkers observed a 91% decline in the formation of $H_2O_2$ when they formed microdroplets in a $N_2$ gas environment[30]. We surmise that the remaining 9% signal was due to the $O_2$ dissolved in their solutions (~4 mg/l at normal temperature and pressure)[33].

When $O_2$ was present in the water, then the ensuing microdroplets contained $H_2O_2$, and the concentration varied with the halides as: $Br^-$ > $Cl^-$ > DI water > $I^-$ (Figure 1-c & S2). While the trend is aligned with the observation of George & co-workers, we doubted that it was due to their charge separation hypothesis[30]. Therefore, we broadened our study to investigate $O_2$ reduction and solid surface oxidation in the presence of halides.

**Effects of halides on the spontaneous formation of $H_2O_2$ at solid–water interfaces**

We chose a common aluminium alloy (AA5083) as the representative material to probe the effects of halides on the spontaneous formation of $H_2O_2$ at solid–liquid interfaces. Freshly polished aluminium specimens of known surface area (~2.3 $cm^2$) were immersed in ~2 ml volume of salt solutions (Figure 2a), equilibrated with air inside sealed glass vials, and the $H_2O_2$ concentrations were measured after 1 hr (with HPAK and cross-checked by $^1$H-NMR). For all the halides, $H_2O_2$ was formed in bulk water in contact with Al plate (Figure 2b). These results, as well as our prior work with a number of metal surfaces such as Mg, Al, and Si, in the presence or absence of $O_2$, reveal that $H_2O_2$ is formed during the reduction of $O_2$ at the solid–liquid interfaces[14]. Figure 2c sheds light on the mechanism –the solid surface oxidizes to soluble products such as metal ions ($M^{n+}$) and insoluble products such as oxides ($MO_x$) and hydroxides ($M(OH)_y$). The insoluble oxidation products form a protective passive layer on the material surface and slow down the oxidation process.

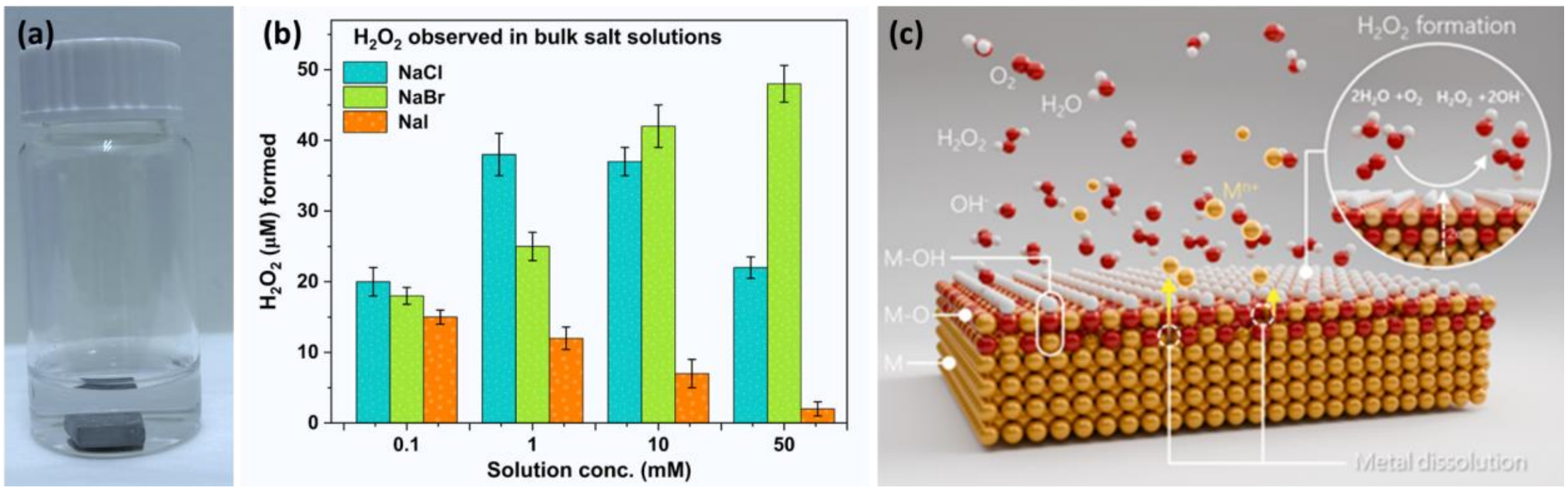

Figure 2 Effects of 0–50 mM $Na^+X^-$ salts, where X: Cl, Br, or I, on the spontaneous $H_2O_2$ formation at the solid–liquid interface in the presence of $O_2$. (a) Freshly polished aluminium specimens were immersed in salt solutions that were equilibrated with air. (b) After 1hr of immersion, the $H_2O_2$ formation at the Al–solution interface for halide concentrations ≤1 mM, followed the trend: $Cl^-$ > $Br^-$>$I^-$, and (ii) For halide concentrations ≥10 mM, the H2O2 formation follows the trend: $Br^-$ > $Cl^-$>$I^-$. (c) Illustration showing the oxidation of the metal surface at the solid–water interface and the reduction of $O_2$ to produce $H_2O_2$. N.B.: Al–water interfacial area of 2.3 $cm^2$ in 10 mM bulk solutions (2 ml).

Curiously, two trends appeared depending on the halides' concentrations (Figure 2b): (i) for salt concentrations ≤1 mM, the $H_2O_2$ formation followed the order: $Cl^-$ > $Br^-$>$I^-$, and (ii) for salt concentrations ≥10 mM, the $H_2O_2$ formation followed the order: $Br^-$ > $Cl^-$>$I^-$. Clearly, the mechanism based on $Br^-$ ions' speciation to the air–water interface, followed by charge separation, fails to explain this observation. Thus, we wondered if the halides could influence the oxidation rates of solid surfaces and be implicated in $H_2O_2$ formation.

**Quantifying Oxidation Rates of Solid Surfaces in the Presence of Salts**

To answer the question raised above, we probed the effects of halides on $O_2$ reduction at the solid–water interface. Aluminum alloy (AA5083) was chosen as the representative material because its oxidation rate can be accurately captured via electrochemical measurements. To gain insights into the surface oxidation (or corrosion), we utilized a potentiodynamic polarization technique (Methods). Typical experiments entailed a three-electrode electrochemical cell with a freshly polished substrate (~1 $cm^2$) as the working electrode, a platinum plate (~1 $cm^2$) as a counter-electrode, and a reference saturated calomel electrode (Methods). To quantify the oxidation rate, first, the open circuit potential (OCP) was measured and then, the potentiodynamic polarization test was commenced at -500 mV below the OCP, scanning a range up to OCP +1000 mV at the rate of +0.5 mV $s^{-1}$. Current densities from these measurements yielded the oxidation rate (or the corrosion rate) due to $O_2$ and the the specific roles of the salts. Figure 3a&b. We found that the surface oxidation rates increased with the salt concentration and always varied with the halides as: $Cl^- > Br^- > I^-$.

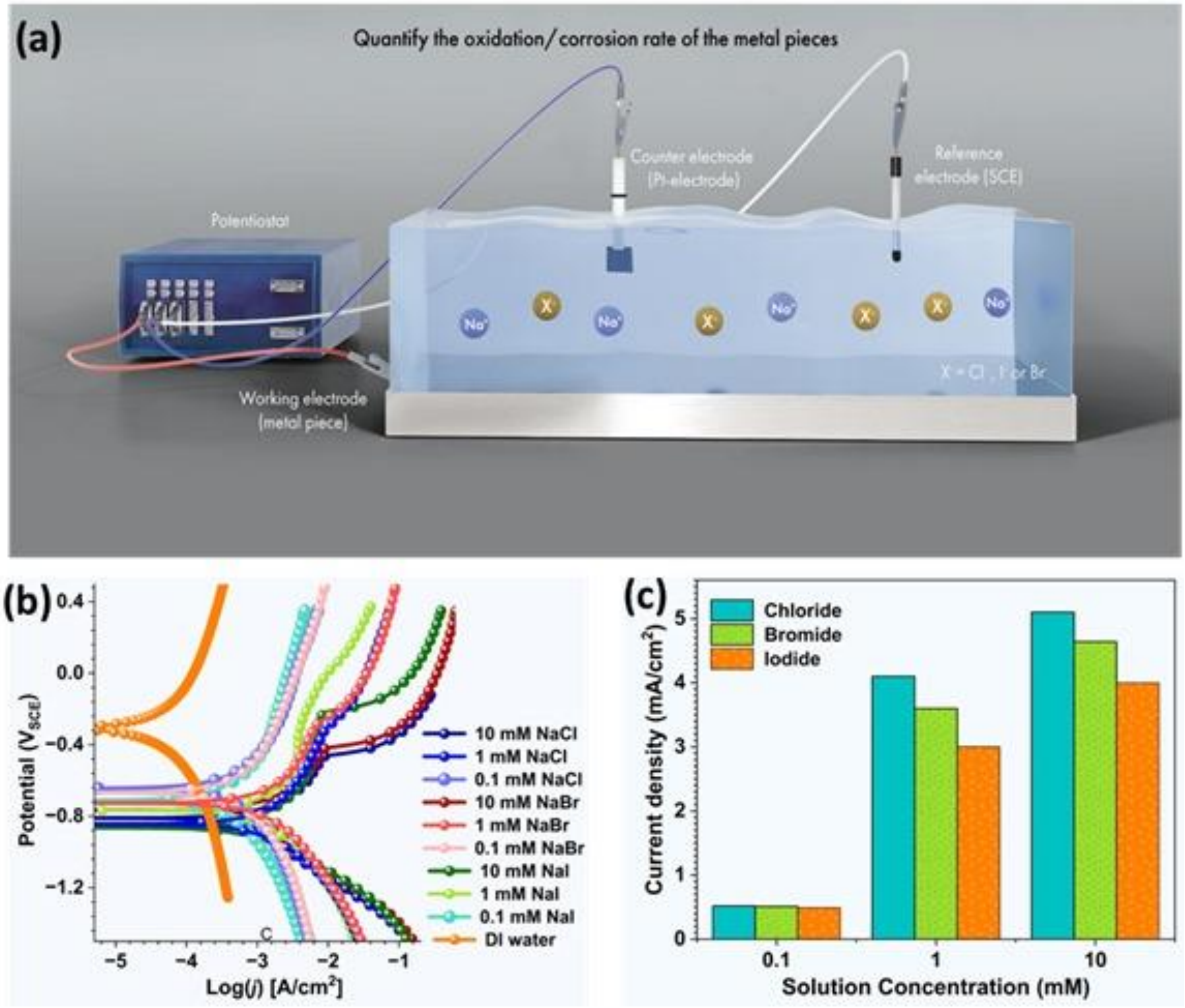

Figure 3. The effect of salts on the oxidation rate of solid/Al at SWI. **(a)** Illustration showing the electrochemical testing setup. It shows a three-electrode cell that consists of a metal surface under investigation acting as the working electrode, a platinum electrode as the counter electrode, and a saturated calomel electrode as the reference electrode. **(b)** Potentiodynamic polarization curves showing oxidation/corrosion behaviors of solid/Al surface in different halide anion solutions. **(c)** Corrosion current densities of the Al (AA5083) surface in different salt solutions calculated from polarization curves. It shows that the oxidation rate of Al in different halide solutions follows the trend $Cl^- > Br^- > I^-$.

Several consistencies and inconsistencies arise as we compare the trends in measured $H_2O_2$ concentrations in sprays (Figure 2b) and the corrosion at the solid–water interface (Figures 3b-c). Firstly, the observed oxidation rates of the metal surface (Figure 3c) follow the same trends as those of the $H_2O_2$ formation at the metal–water interface for salt concentrations ≤1 mM, i.e., $Cl^- > Br^- > I^-$ (Figure 2b). However, for salt concentrations ≥10 mM, the $H_2O_2$ formation follows a different order: $Br^- > Cl^- > I^-$ (Figure 2b). Next, for $I^-$, whereas the oxidation rates keep on increasing with the concentration, the measured $H_2O_2$ concentrations follow the opposite trend. How do we reconcile these findings?

**The Curious Case of $Cl^-$ and $Br^-$ vis-à-vis Surface Oxidation**

To understand why the $H_2O_2$ formation at the solid–water interface decreases with the $Cl^-$ concentration ≥10 mM, one has to dive into the fundamentals of pitting corrosion[35]. It refers to the formation of a pit across the passive oxide layer, exposing the fresh metal surface and restarting corrosion. Extensive scientific literature reveals that the halide ions' ability to cause pitting follows this trend: $F^- > Cl^- >> Br^- > I^-$.[36, 37] Pitting depends on the material characteristics and the halide concentration. For instance, for the aluminum alloy chosen in this work, pitting becomes aggressive at ≥10 mM (Figure S4a). So, while pitting is insignificant at 1 mM concentration, it becomes severe at 10 mM, exposing a fresh metal surface. As the fresh metal surface comes into contact with $H_2O_2$, oxidation follows. As shown in Figure S4-a, the onset of pitting is marked by a sharp rise in the current density. Interestingly, $H_2O_2$ that was formed initially as the byproduct of $O_2$ reduction at the solid–water interface is now consumed during pitting. As pitting by $Cl^-$ ions continues, and this process repeats, the $H_2O_2$ concentration keeps decreasing. This reconciles the concentration-dependent trends in the measured $H_2O_2$ concentration above and the measured current densities (Figures 2b & 3c), and calls into question the explanation put forth by George & co-workers[30].

**The Curious Case of $I^-$ vis-à-vis Surface Oxidation**

The formation of $H_2O_2$ at the solid–water interface decreases with increasing NaI concentration, as presented in Fig. 2b and also noted by Georged & co-workers[30]. This can be explained based on the reaction of $H_2O_2$ with iodide ions, leading to the formation of triiodide ($I_3^-$). See Figure S3, where the evidence for the formation of $I_3^-$ is presented.

**Reconciling $H_2O_2$ formation and $O_2$ Reduction at the Solid–Water Interfaces**

Based on the experimental results (Figs. 2, 3, S4, and S7), it is clear that the peculiar properties of halides vis-à-vis pitting and reactivity with $H_2O_2$ lead to their consumption over time. To prove that the $H_2O_2$ formation is directly proportional to the oxidation of solid surfaces, we devised a new experiment. We exploited the fast reactivity of the hydrogen peroxide assay kit (HPAK) reaction mixture with $H_2O_2$, leading to a fluorescent product, to preempt its consumption by pitting or $I^-$. Metal (Al) specimens were immersed in a 1:1 volumetric mixture comprised of HPAK kit and 10 mM salt solutions and the fluorescence was measured. The results reveal that the $H_2O_2$ formation at the solid–water interface followed the same order as that of the oxidation rate of the solid surface, i.e., $Cl^- > Br^- > I^-$ (Figure 4).

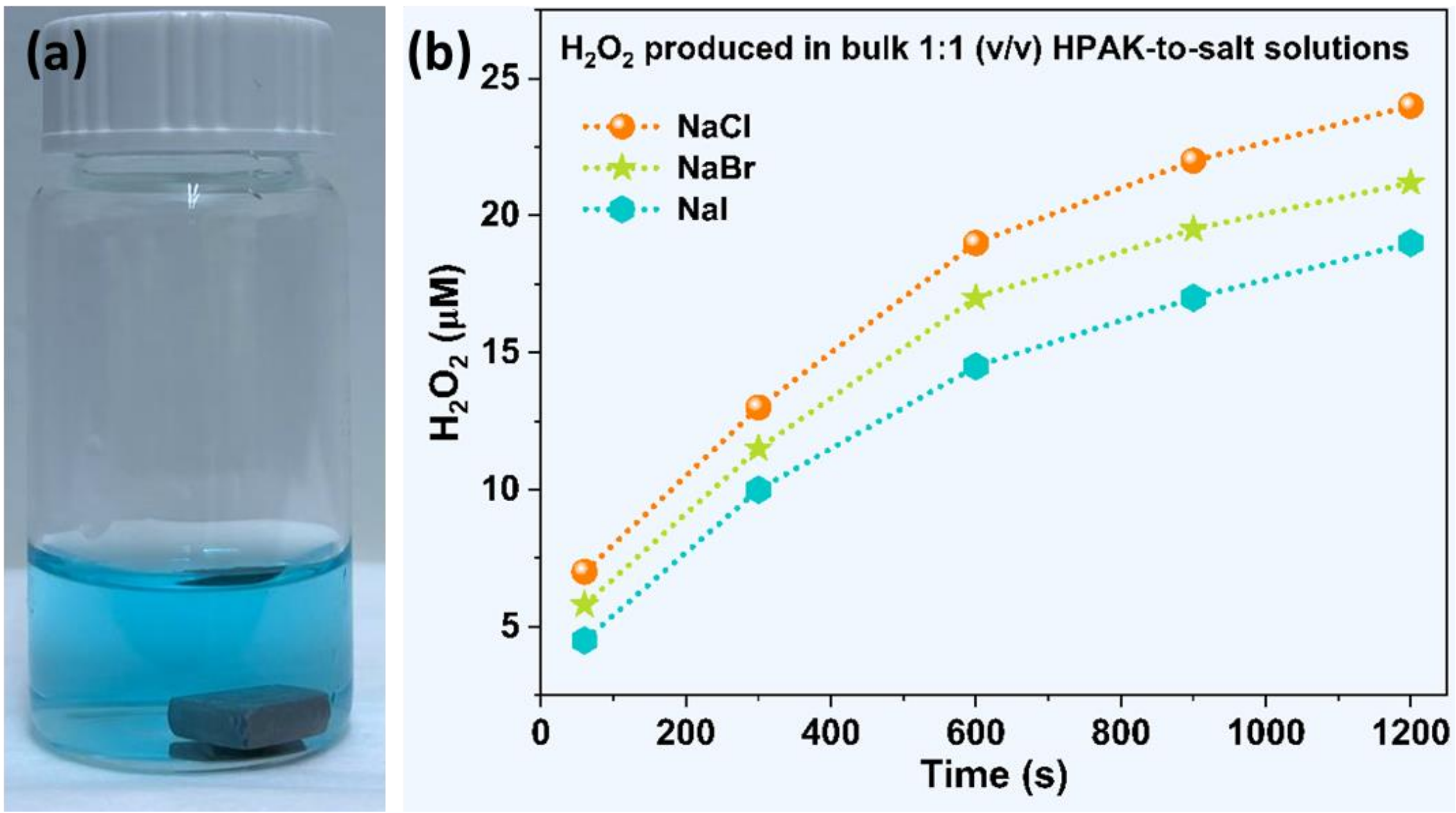

Figure 4 Time-dependent measurement of $H_2O_2$ formation at the solid–water interface in solutions obtained by mixing 10 mM $Na^+X^-$ solutions, where X: Cl, Br, or I, with HPAK reaction mixture (1:1 v/v). See details of the measurements in the Methods section.

In summary, we noted that $H_2O_2$ formation is directly related to the oxidation of the solid surface: no $O_2$, no $H_2O_2$. The $H_2O_2$ formation is highest in NaCl solution, followed by NaBr, and least in NaI solution, having the same salt concentrations, their tendency to accelerate the corrosion or oxidation of the solid/Al surface follows the same order. Our findings explain the dataset of George & coworkers[30] in great detail and refute their mechanism that suggests that $Br^-$ drives the $H_2O_2$ formation at the air–water interface due to charge separation.

**Effect of water pH on the formation of $H_2O_2$ at the solid–water interfaces**

Lastly, we examined the claims of George & co-workers[31] for the pH-dependence of $H_2O_2$ formation, i.e., the higher the alkalinity, the higher the $H_2O_2$ formation. We apply our framework of finding the equivalence between $O_2$ reduction and solid surface oxidation vis-à-vis $H_2O_2$ formation. To do this, we chose aluminum and titanium as representative materials due to their contrasting pH-dependent oxidation behavior, i.e., Ti is known to passivate more (less oxidation) in acidic solutions, while Al passivates more in alkaline solutions. This way, we could disentangle the contributions of the solution alkalinity versus the surface oxidation towards $H_2O_2$ formation.

Specimens of the Al alloy (AA5083) with a surface area of 2.3 $cm^2$ were immersed in the following aqueous solutions (2 ml volume) equilibrated with air (i.e., containing $O_2$) for 1 hour: water (pH 5.6), pH 2 water (adjusted with HCl), and pH 10 (adjusted with NaOH). HPAK measurements (and cross-validation by 1H-NMR) revealed that for the Al–water interface, the $H_2O_2$ formation followed the following trend: pH 2 (~47 μM) > pH 10 (~14 μM) > water (~4.5 μM) (Figure 5a). In contrast, the Ti-water interface exhibited the following trend for the $H_2O_2$ formation: pH 10 (~0.88 μM) > pH 2 (~0.42 μM) > water (~0.22 μM) in one hour (Figure 5b). These results demonstrate that the previously proposed mechanisms for the pH as a driver for enhancing $OH^-$ ions at the air–water interface, leading to the $H_2O_2$ formation, are wrong[32 31]. Instead, it is the nature of the solid surface that gets oxidized that governs the pH-dependence of the $H_2O_2$ formation. To elucidate the last point further, we utilized the electrochemical potentiodynamic polarization technique to quantify the oxidation behaviors of Al and Ti specimens immersed in solutions of acidic/alkaline/neutral pH for 1 hour. The experimental results revealed that the oxidation rates exhibited the same trends as the $H_2O_2$ formation vis-à-vis the solution pH (Figures 5c–d). Specifically, the corrosion/oxidation rate ($i_{corr}$) of Al followed the order: $i_{corr}$ acidic solution (~118.5 mA/$cm^2$) > $i_{corr}$ alkaline solution (~2.8 mA/$cm^2$) > $i_{corr}$ DI water (~79.4 μA/$cm^2$) (Figure 5c). For Ti, the corrosion rate ($i_{corr}$) followed the order $i_{corr}$ alkaline solution (~2.8 mA/$cm^2$) > $i_{corr}$ acidic solution (~0.45 mA/$cm^2$) > $i_{corr}$ DI water (~0.063 mA/$cm^2$) (Figure 5d).

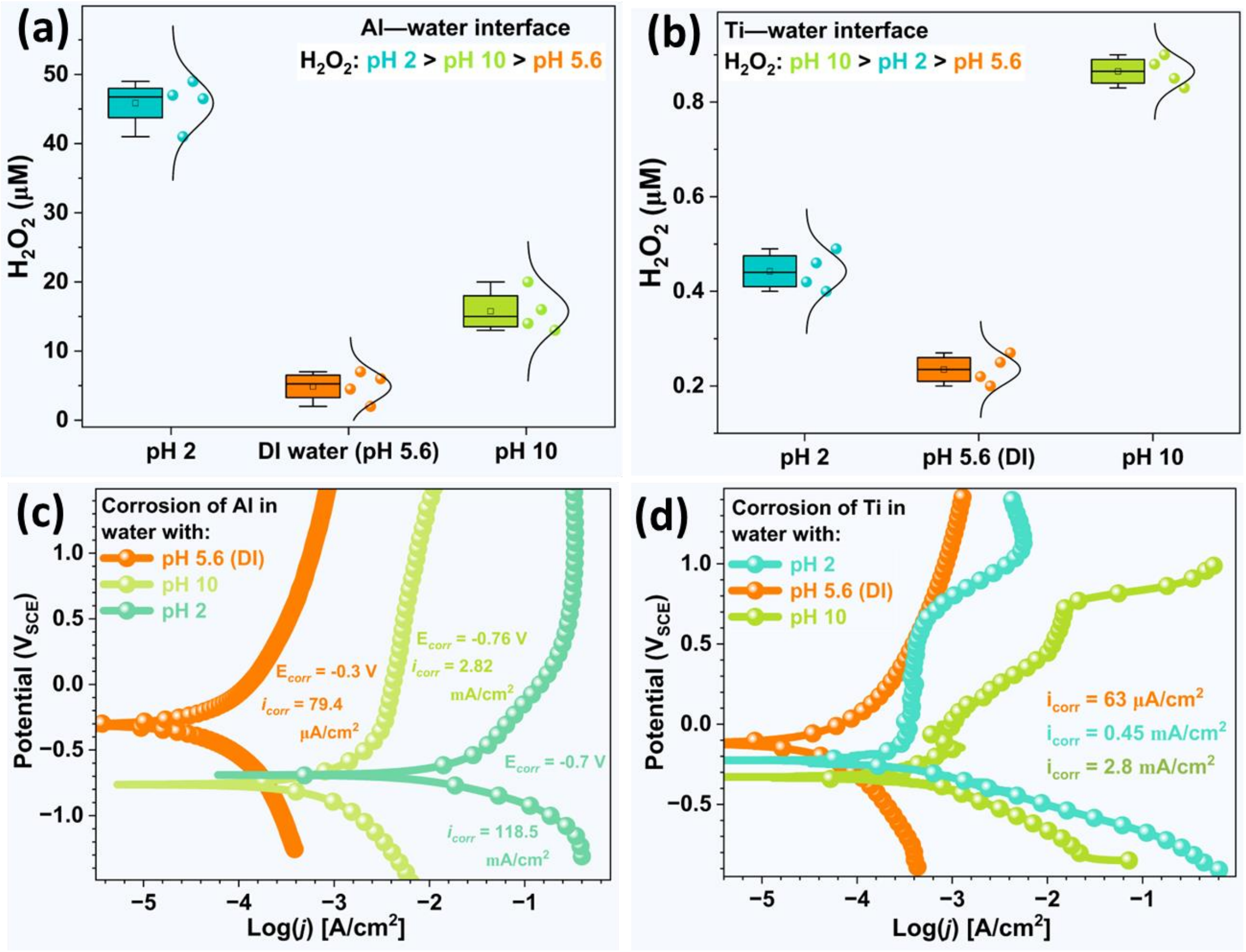

Figure 5 Effect of pH on $H_2O_2$ formation at water–solid interface. **(a)** Effect of pH on the $H_2O_2$ produced at the Al–water interface. **(b)** Effect of pH on the $H_2O_2$ produced at the Ti–water interface. **(c)** Effect of pH on the oxidation or corrosion behavior of the Al–water interface. **(d)** Effect of pH on the oxidation or corrosion behavior of the Ti–water interface.

In summary, our experimental results on the effect of halide anions or solution pH on $H_2O_2$ formation revealed that $H_2O_2$ is directly related to the oxidation of solid material at the solid–water interface. The solution conditions that lead to more corrosion of the solid surface led to more $H_2O_2$ formation. While we explore the oxidation of two metallic materials (Al & Ti) in this paper, other non-metallic materials (like glass) also get corroded in aqueous solutions and produce $H_2O_2$.[14, 38, 39]

## Conclusion

Taken together, our experimental results demonstrate that the air–water interface or the microdroplet geometry is incapable of spontaneously forming $H_2O_2$ regardless of the halides or the presence/absence of $O_2$. We have pinpointed solid–water interfaces as the sites for the spontaneous $H_2O_2$ formation – only in the presence of $O_2$, i.e., no $O_2$, no $H_2O_2$. In this process, $O_2$ is reduced, and the solid surface is oxidized (i.e., corrosion). While solid–water interfaces can be conveniently bypassed in computer simulations, it is not possible inside a physical laboratory. Thus, we refute the previously reported mechanisms and conclusions, identifying the air–water interface as the site for spontaneous $H_2O_2$ formation[1, 2, 4, 8, 10-12, 29-31].

## Methods:

### Chemicals.

The following chemicals were purchased from Sigma Aldrich: sodium chloride (NaCl, 7647145), sodium bromide (CAS #7758-02-3), standard hydrogen peroxide ($H_2O_2$) 30% (Cat.270733), HPLC grade water (Cat. 2594649). Sodium iodide was obtained from Ward's Science (CAS #7681825). Water was obtained from a MilliQ Advantage 10 set up (with resistivity 18.2 MΩ cm) and referred to as "deionized" (DI). The HPAK (fluorometric-near infrared) kit was purchased from Abcam® (ab138886).

### Quantification of $H_2O_2$

The concentration of $H_2O_2$ in all the diluted salt solutions was determined utilizing the hydrogen peroxide assay kit (HPAK) and cross-validated by NMR (described below). The HPAK method relies on the interaction between hydrogen peroxide and the AbIR peroxidase indicator, leading to fluorescence emission. The maximum excitation and emission wavelengths for this reaction are 647 nm and 674 nm, respectively. To analyze the samples, a mixture of 50 μL of HPAK reaction solution and 50 μL of the samples was prepared in a 96-well black/transparent bottom microtiter plate and read using a SpectraMax M3 microplate reader. The fluorescence data were processed using the SoftMax Pro 7 software. The $H_2O_2$ content in the samples was quantified by referencing the calibration curve established with standard samples on the same day.

### NMR spectroscopy for $H_2O_2(aq)$ quantification

Nuclear Magnetic Resonance (NMR) measurements were carried out utilizing the 950 MHz Bruker Avance Neo NMR spectrometer, which was equipped with a 5-mm Z-axis gradient TXO Cryoprobe and operated at 275 K. In all experiments, a 6-ms Gaussian 90-degree pulse was utilized to excite the proton of hydrogen peroxide, followed by a 50-ms data acquisition period and a 1-ms recycling delay. The quantification of peak intensities and the verification of peak positions corresponding to $H_2O_2$ in salt solutions were conducted by comparing the results obtained from standard $H_2O_2$ samples. A total of 50000 scans were recorded in each case, which enables a detection of $H_2O_2$ as low as 50 nM.

## Electrochemical testing

The corrosion behavior of solid specimens (Al, Ti) in aqueous solutions was assessed through potentiodynamic polarization tests. Right before each experiment, the specimens were meticulously polished using SiC grit papers up to 1200 grit, followed by cleansing with high-pressure $N_2$ gas. All experiments were conducted on specimens with a surface area of 1 $cm^2$, utilizing a BioLogic VMP3 potentiostat. A three-electrode system was consistently employed, with a platinum sheet (1 $cm^2$) serving as the counter electrode and a saturated calomel electrode (SCE) as the reference electrode. The tests were carried out at a scan rate of 0.5 mV $s^{-1}$ and were performed. The open circuit potential (OCP) was only monitored for 5 seconds in each instance to prevent or minimize corrosion during the stabilization phase. The potentiodynamic polarization tests commenced at -500 mV with respect to the OCP.

## Acknowledgments

The co-authors thank Dr. Adair Gallo and Ms. Nayara Musskopf for building a robust experimental setup in HM's laboratory; Dr. Ana Teresa Bigio, Scientific Illustrator, for the graphical abstract (TOC), Figure 1a, and Figure S1; Dr. Ryo Mizuta, Scientific Illustrator, for preparing illustrations Fig. 2b; Dr. Sankara Arunachalam and Dr. Krishna Katuri for helping with SEM imaging and providing a potentiostat for electrochemical measurements, respectively; Prof. Willem Koppenol (ETH Zurich) for fruitful discussions.

**Author contributions:** ME designed and performed the experiments and analyzed the data. HM oversaw the research and secured funding. ME and HM wrote the manuscript.

**Competing interests:** The authors declare no competing interests.

**Data and materials availability:** All data needed to evaluate the conclusions in the paper are present in the paper or the Supplementary Materials.

**Funding:** HM acknowledges KAUST for funding (Grant No. BAS/1/1070-01-01).

# Supplementary Information

# Disentangling the Roles of Dissolved Oxygen, Common Salts, and pH on the Spontaneous Hydrogen Peroxide Production in Water: No $O_2$, No $H_2O_2$

Muzzamil Ahmad Eatoo[1,2,*] & Himanshu Mishra[1,2,*]

[1] Environmental Science and Engineering (EnSE) Program, Biological and Environmental Science and Engineering (BESE) Division, King Abdullah University of Science and Technology (KAUST), Thuwal, 23955-6900, Kingdom of Saudi Arabia

[2] Interfacial Lab (iLab), King Abdullah University of Science and Technology (KAUST), Thuwal 23955-6900, Saudi Arabia

*Correspondence: muzzamil.eatoo@kaust.edu.sa and himanshu.mishra@kaust.edu.sa

## Section S1. Spray setup for microdroplet formation

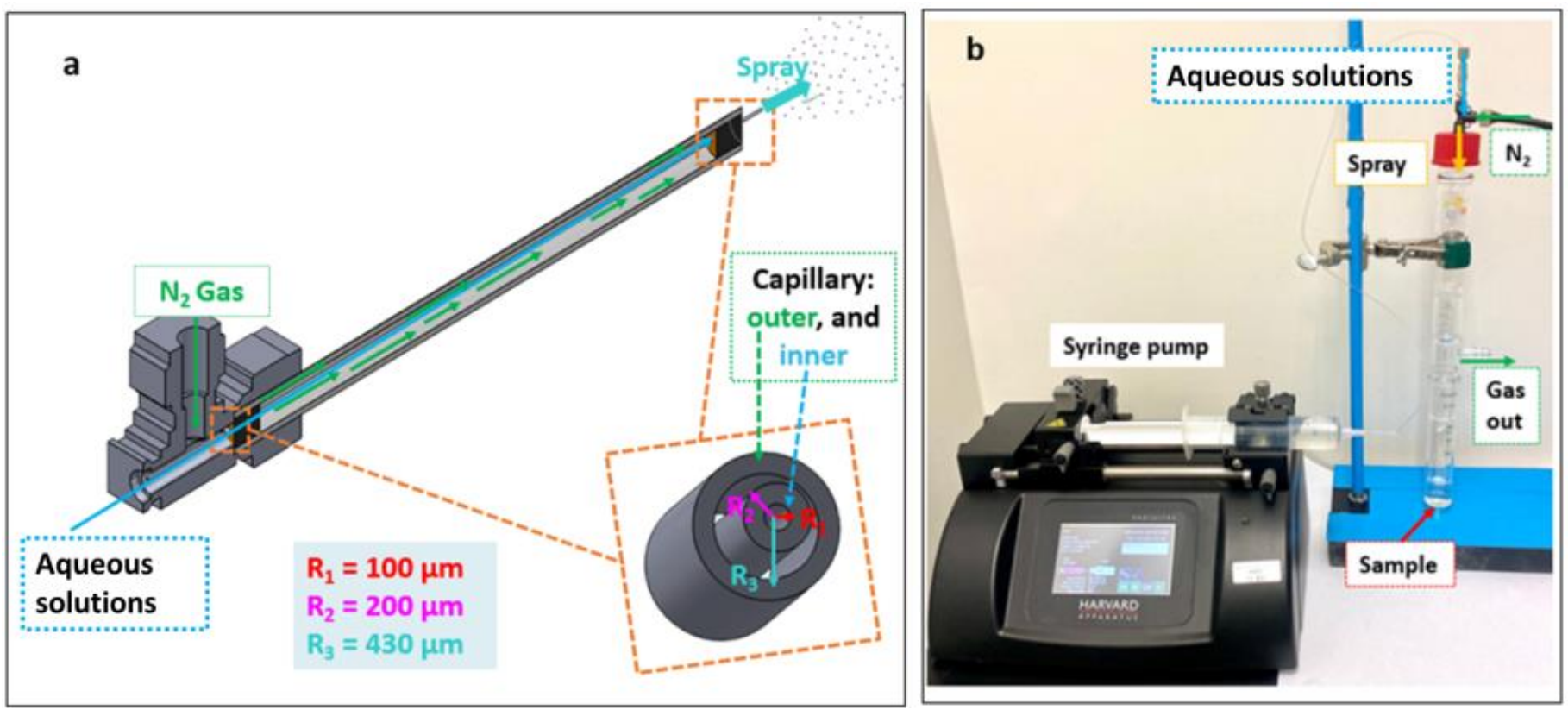

Fig. S1: (a) Schematics of pneumatic spray setup. (b) Photograph of a spray setup featuring a detachable glass flask for sample collection, incorporating two coaxial capillary tubes. The liquid sample is delivered through the inner capillary, while nitrogen gas flows through the surrounding outer annulus. Halide salt solutions were introduced using a syringe pump, and nitrogen gas was supplied from a high-pressure cylinder. The resulting microdroplets were sprayed and collected in a clean glass flask until an adequate volume of analyte had been obtained.[1, 2]

## Section S2. Chemicals and Experimental Methods

S2.1 Chemicals: The following chemicals were purchased from Sigma Aldrich: sodium chloride (NaCl, 7647145), sodium bromide (CAS #7758-02-3), standard hydrogen peroxide ($H_2O_2$) 30% (Cat.270733), HPLC grade water (Cat. 2594649). Sodium iodide was obtained from Ward's Science (CAS #7681825). Water was obtained from a MilliQ Advantage 10 set up (with resistivity 18.2 MΩ cm) and referred to as "deionized" (DI). The HPAK (fluorometric-near infrared) kit was purchased from Abcam® (ab138886).

**S2.2 Measurement of $H_2O_2$ formation in microdroplets:** The microdroplets were formed by spraying bulk solutions at a flow rate of 25 μL/min using high-pressure (100 psi) dry $N_2$ gas. After spraying, the microdroplets were collected, and $H_2O_2$ was measured using $^1$H-NMR. Nuclear Magnetic Resonance (NMR) measurements were carried out utilizing the 950 MHz Bruker Avance Neo NMR spectrometer, which was equipped with a 5-mm Z-axis gradient TXO Cryoprobe and operated at 275 K. In all experiments, a 6-ms Gaussian 90-degree pulse was utilized to excite the proton of hydrogen peroxide, followed by a 50-ms data acquisition period and a 1-ms recycling delay. Over 50000 scans were collected with a recycle delay of 1 ms between scans[1].

**S2.3 Examination of $H_2O_2$ formation in bulk solution**. The $H_2O_2$ formation at the solid-water interface in bulk solutions was examined using Metal samples (Al or Ti). Metal specimens of the same total surface area were submerged in different solutions of equal volumes, and the $H_2O_2$ formed was measured using $^1$H-NMR and HPAK methods.

**S2.4 Investigation of the oxidation rate of the solid surface.**

The corrosion behavior of solid specimens (Al, Ti) in aqueous solutions was assessed through potentiodynamic polarization tests. Right before each experiment, the specimens were meticulously polished using SiC grit papers up to 1200 grit, followed by cleansing with high-pressure $N_2$ gas. All experiments were conducted on specimens with a surface area of 1 $cm^2$, utilizing a BioLogic VMP3 potentiostat. A three-electrode system was consistently employed, with a platinum sheet (1 $cm^2$) serving as the counter electrode and a saturated calomel electrode (SCE) as the reference electrode. The tests were carried out at a scan rate of 0.5 mV $s^{-1}$ and were performed. The open circuit potential (OCP) was only monitored for 5 seconds in each instance to prevent or minimize corrosion during the stabilization phase. The potentiodynamic polarization tests commenced at -500 mV with respect to the OCP.[3]

**S2.5. Quantification of $H_2O_2$ in sprayed water microdroplets by HPAK kit.**

The concentration of $H_2O_2$ in all the diluted salt solutions was determined utilizing the hydrogen peroxide assay kit (HPAK). This method relies on the interaction between hydrogen peroxide and the AbIR peroxidase indicator, leading to fluorescence emission. The maximum excitation and emission wavelengths for this reaction are 647 nm and 674 nm, respectively. To analyze the samples, a mixture of 50 µL of HPAK reaction solution and 50 µL of the samples was prepared in a 96-well black/transparent bottom microtiter plate and read using a SpectraMax M3 microplate reader. The fluorescence data were processed using the SoftMax Pro 7 software. The $H_2O_2$ content in the samples was quantified by referencing the calibration curve established with standard samples on the same day

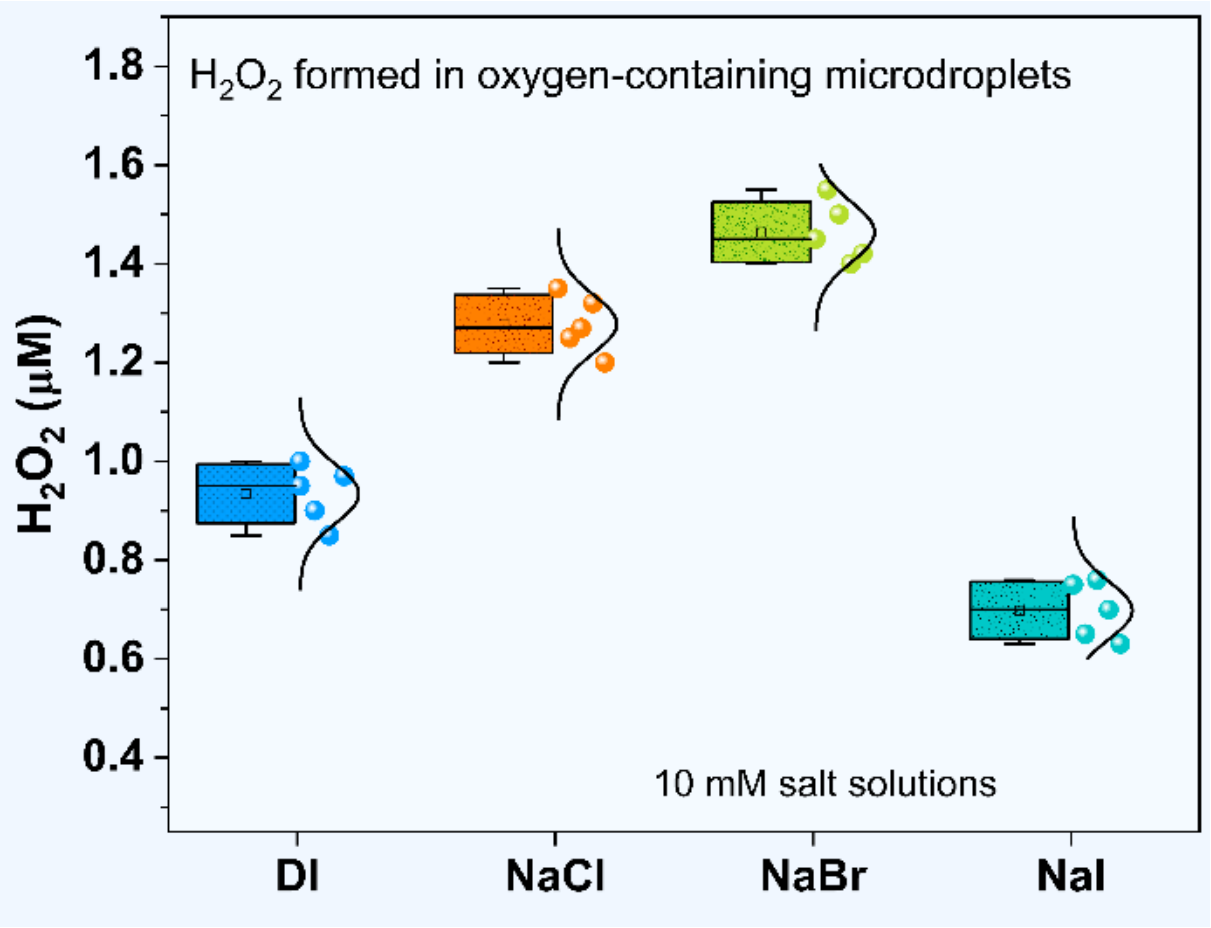

Figure S2. Quantification of $H_2O_2$ formed in sprayed microdroplets by HPAK.

## Section S3. Oxidation of iodide ions to triiodide by $H_2O_2$

Oxidation of iodide ions to triiodide due to $H_2O_2$ produced by the Al specimen in 10 mM NaI solution in 1 day. The oxidation of iodide ions by $H_2O_2$ produced at the Al-water interface due to $O_2$ reduction, to triiodide, was examined by UV-vis spectroscopy. Blank 10mM NaI without metal specimen was used as a reference sample. And the 10 mM NaI sample with a freshly polished Al specimen for 24 hours was scanned in the range of 200-600 nm using quartz cuvettes with a 1 cm path length. The triiodide ion ($I_3^-$) typically shows a characteristic absorption peak near 350 nm, and a weaker band around 290 nm was observed, as shown in Figure S4 below.

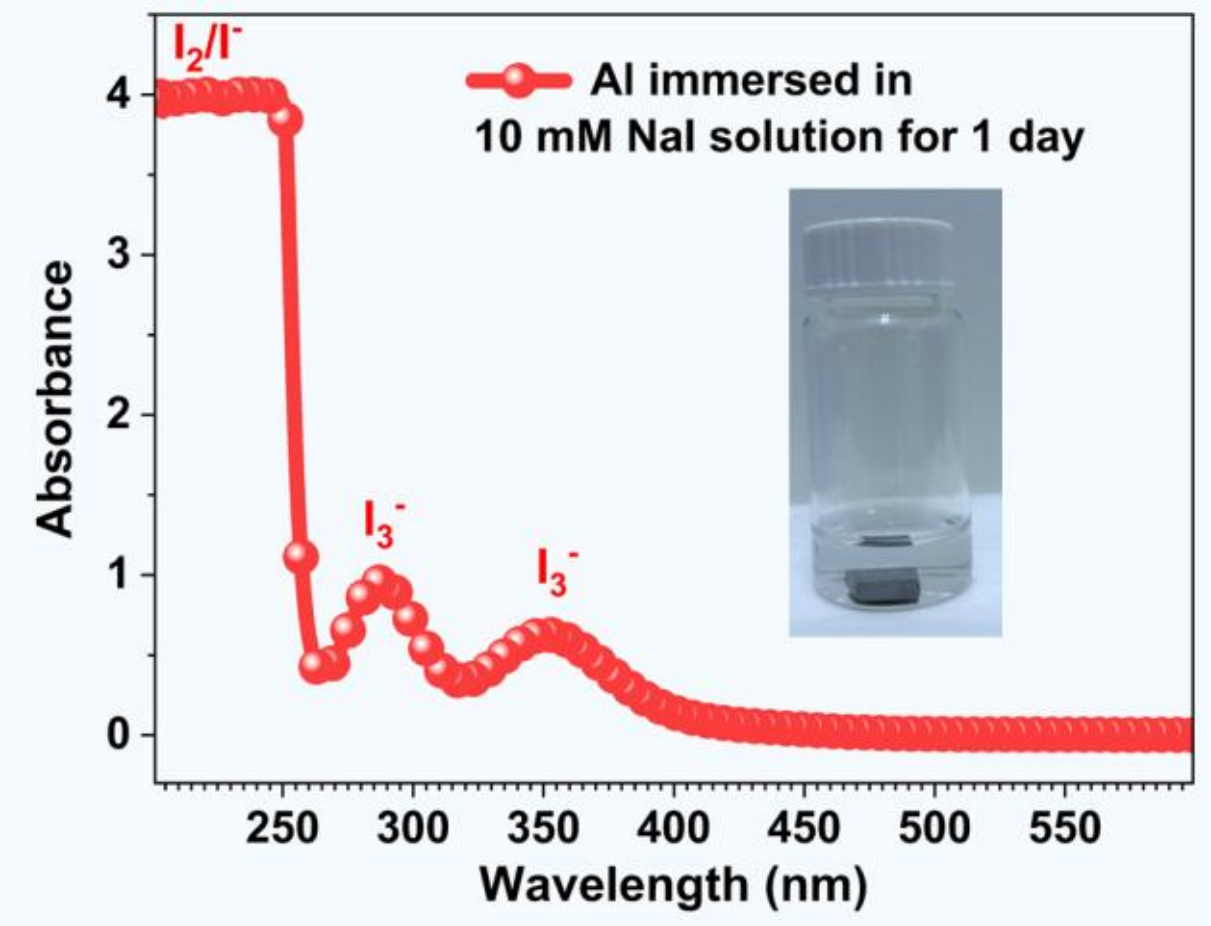

**Fig. S3.** shows the formation of triiodide in NaI solution due to the presence of $H_2O_2$ formed by the Al sample in the 10 mM solution.

## Section S4. Pitting corrosion by Cl⁻ ions

We investigated the Al surfaces after exposure to salty solutions via scanning electron microscopy (SEM) (Fig. S5) and found and paid closer attention to the polarization curves given in Fig. 4a. It is evident from both polarization curves and micrographic images that the Al surface undergoes severe pitting attack at higher NaCl concentrations. In Fig. 4a, the pitting is represented by a sudden increase in anodic currents. It can be clearly observed that 10 mM and 50 mM salt concentrations cause severe pitting. The pits formed in the solid Al surface can also be clearly seen in SEM micrographs (Fig. S5).

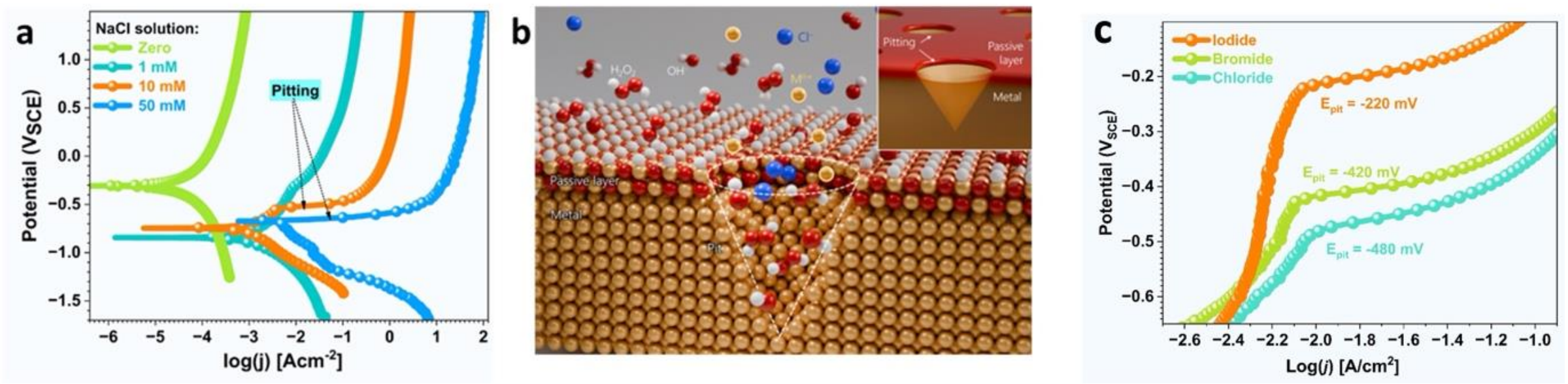

**Fig. S4** The effect of Cl⁻ concentration on corrosion behavior of Al (AA5083) in NaCl solutions. **a)** Potentiodynamic polarization curves showing pitting of Al in 10 mM and 50 mM NaCl solutions. **b)** Illustration showing the pitting or passive layer breakdown phenomenon by $Cl^-$ ions. **(c)** The effectiveness of $Cl^-$, $Br^-$, and I⁻ ions to cause pitting. The data shows the pitting potentials of Al in 10 mM solutions of NaCl, NaBr, and NaI.

The pitting tendency of different halide anions was also examined by measuring the pitting potentials. The pitting potentials of Al in 10 mM salt solutions of NaCl, NaBr, and NaI are around -480, -420, and -220 mV, respectively (Fig. S4-c). This shows that among different halide anions, chloride has the highest pitting power. Therefore, the pitting power of halide ions follows the trend $Cl^- > Br^- > I^-$.

**Section S5.** SEM imaging of Al after exposure to NaCl solution.

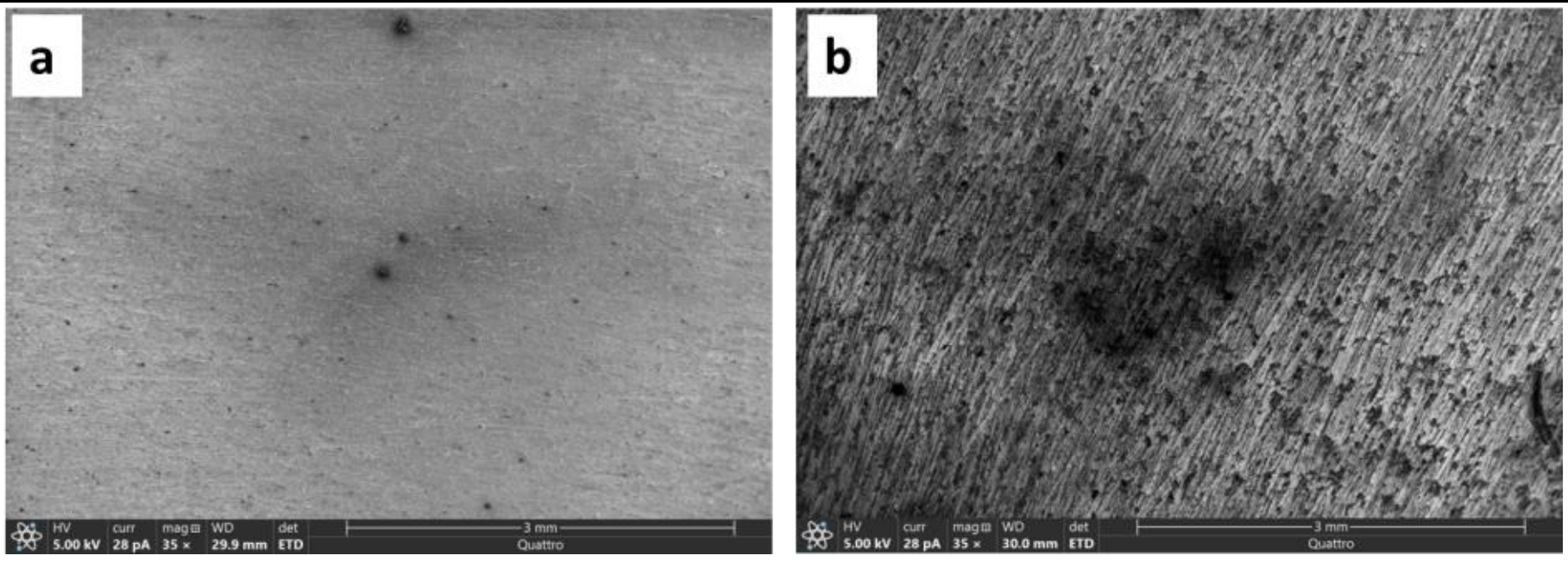

Fig. S5. SEM images of Al immersed for 24 hours in: a) 10 mM NaCl solution. b) 50 mM NaCl solution. The pitting attack represented by pore-like features in the images can be clearly observed. Also, in 50 mM NaCl solution, the high pit density and damage of the oxide layer on the surface are evident. This pitting or film breakdown brings the naked metal surface in direct contact with $H_2O_2$ present in the solution. After that, the $H_2O_2$ gets consumed by the metal surface.

**Section S6:** To examine the effect of pitting or passive film breakdown on $H_2O_2$ present in the solution, we investigated the interaction of $H_2O_2$ with the fresh metal surface (polished) and thermally grown oxide on the metal/Al surface. We prepared a 100 μM $H_2O_2$ solution using deaerated water so that the Al samples do not produce any $H_2O_2$. We immersed Al samples having the same surface area (2.3 $cm^2$) in equal volumes of 100 μM $H_2O_2$ solutions. One of the samples was freshly polished to remove any impurity or oxide on its surface, and an oxide layer was thermally grown on the other sample by heating it at ~500 °C in a muffle furnace for 12 hours in the presence of $O_2$. The results obtained after 1 hour of immersion in $H_2O_2$ solution are shown in Fig. S6. The results revealed that $H_2O_2$ is consumed more by the fresh metal surface as compared to one covered with an oxide layer.

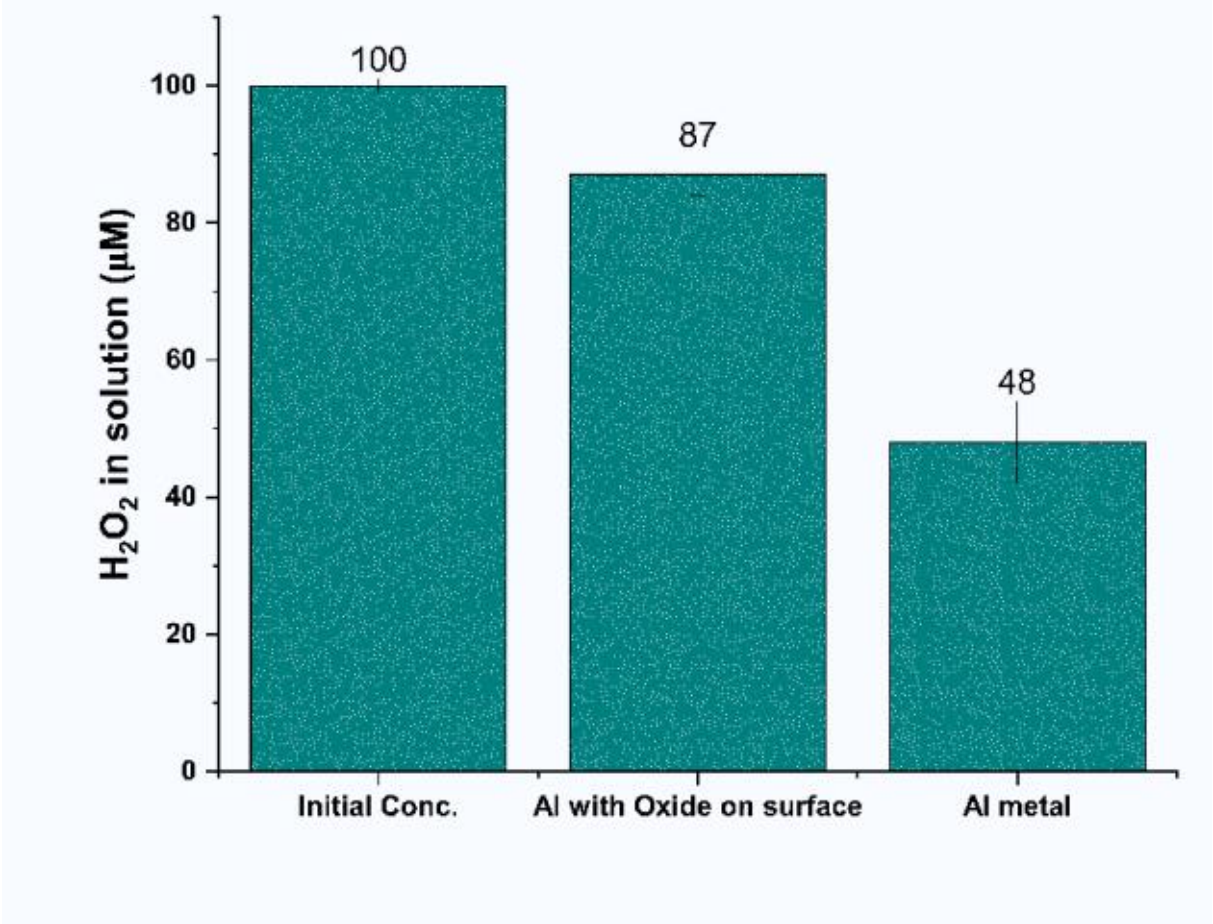

**Fig. S6** Comparison of $H_2O_2$ consumption by fresh metal surface vs the surface with oxide layer.